# Why Can't We Predict RNA Structure At Atomic Resolution?

Kyle Beauchamp[1, ‡], Parin Sripakdeevong[1, ‡], Rhiju Das[1,2,*]

kyleb@stanford.edu , sripakpa@stanford.edu , rhiju@stanford.edu

(1) Biophysics Program, Stanford University (2) Biochemistry Department, Stanford University

‡ Equally contributing authors.  *Corresponding author

## 4.0 Abstract

No existing algorithm can start with arbitrary RNA sequences and return the precise, three-dimensional structures that ensures their biological function.  This chapter outlines current algorithms for automated RNA structure prediction (including our own FARNA-FARFAR), highlights their successes, and dissects their limitations, using a tetraloop and the sarcin/ricin motif as examples.  The barriers to future advances are considered in light of three particular challenges: improving computational sampling, reducing reliance on experimentally solved structures, and avoiding coarse-grained representations of atomic-level interactions.  To help meet these challenges and better understand the current state of the field, we propose an ongoing community-wide CASP-style experiment for evaluating the performance of current structure prediction algorithms.

## 4.1 RNA as a Model System

Predicting the three dimensional structures of biopolymers from their primary sequence remains an unsolved but foundational problem in theoretical biophysics. This problem lies at the frontier of modern biological inquiry, encompassing questions from folding of individual protein and RNA domains to the fiber assembly of histone-compacted DNA genomes. However, a predictive, atomic-resolution understanding of these three-dimensional processes is presently out of reach. Attaining such an understanding will likely require simple starting points, and we view the folding of small RNA systems as the most tractable of these unsolved puzzles.

Beyond validating and refining our physical understanding of biomolecule behavior, a general algorithm to model RNA structure would have immediate practical implications. Riboswitches, ribozymes, and new classes of functional non-coding RNAs are being discovered rapidly, through RNA secondary structure prediction algorithms, bioinformatic tools, and a large suite of experimental approaches. Accurate and fast tools for predicting three dimensional structure would not only accelerate these discoveries but lead to richer experimentally testable hypotheses for how these molecules sense the cellular state and bind recognition partners. Furthermore, accurate three-dimensional RNA models would expand the use of RNA as a designer molecule, with potential applications ranging from the control of organisms [see, e.g., (Win et al. 2009)], the engineering of nano-scaffolds [see, e.g., (Jaeger and Chworos 2006)], the development of aptamer-based therapeutics [see, e.g., (Nimjee et al. 2004)], and the emerging fields of nucleic acid computation and logic [see, e.g., (Stojanovic and Stefanovic 2003)].

This chapter discusses the present state of computational modeling of three-dimensional RNA structure, highlighting successes and describing the barriers to future progress.  Our hope is that dissecting the limitations of the field will hasten the development of methods for modeling RNA structures without extensive experimental input.

**4.2 Is the RNA Structure Prediction Problem well defined?**

RNA, despite its small four-letter alphabet, is now recognized to perform a multitude of roles in the cell, including information transfer, catalysis (Nissen et al. 2000), gene regulation, and ligand sensing (Mandal and Breaker 2004). The attainment of a small set of unique three-dimensional states has been a hallmark of previously characterized functional biomolecules, from catalytic proteins to information-storing DNA double helices. Do RNA molecules of the same type have structures agreeing at atomic resolution, up to the fluctuations expected of a biomolecule in solution?  Is the information necessary to specify these structures contained in the RNA sequence alone?

 We now know that the answer to both questions is "yes" for a broad range of natural and *in vitro* selected RNA sequences, although there are also examples of both unstructured RNAs and RNAs guided into functional conformations by partners (induced fit; see, e.g., (Ferre-D'amare A and Rupert 2002 ; Hainzl et al. 2005)). In the 1960s, studies of transfer RNA sequences defined a conserved secondary structure [see, e.g., (Holley et al. 1965; Shulman et al. 1973 ; Rich and RajBhandary 1976 ) ]– the pattern of classic Watson-Crick base pairs – and then defined inter-helical tertiary interactions mediated by non-canonical base-base contacts [see, e.g., (Levitt 1969 ; Kim et al. 1974)]. These pioneering studies established a paradigm of theoretical investigation and experimental decipherment that has been followed for each novel class of RNAs that has been discovered in subsequent decades. In many respects, it now appears that RNA is easier to fold than other biopolymers. For example, unlike proteins, which typically require at least a dozen residues to form well-defined structures, the simplest RNAs with well-defined, recurrent structures are as small as eight residues (Jucker et al. 1996).

   These simple molecules include hairpin loops, single strands that fold back on themselves to form short Watson-Crick helices. In some cases, the loops contain only four non-helical bases—the so-called tetraloops (Varani 1995), with two classes, UUCG and GCAA (with their respective homologues), being the most extensively studied (see Fig. 1a) (Antao and Tinoco Jr 1992; Jucker and Pardi 1995; Molinaro and Tinoco Jr 1995; Jucker et al. 1996; Correll et al. 2003). These motifs have been observed in isolation, as single strands of RNA (Jucker et al. 1996), and as segments within larger RNA structures. Spectroscopic, crystallographic, and thermodynamic experiments indicate that these tetraloops form stable structures that are largely conserved among homologous sequences.

   Larger RNA systems exhibit well-defined three-dimensional folds as well, and work over the last decade has yielded a rich trove of crystallographic structures of ligand binding aptamers, riboswitches, and ribozymes. Most famously, ribosomal subunits of several organisms have been crystallized by several groups (Ban et al. 2000; Wimberly et al. 2000; Harms et al. 2001; Yusupov et al. 2001), and the resulting structures are remarkably similar.  For example, the conserved structural core shared by the respective 16S and 23S rRNAs of E. coli and T. thermophilus, two bacteria that diverged early in evolution, comprises 90% or more of these molecules, despite extensive sequence differences (Zirbel et al. 2009).  Similar stories of intricate structures shared across homologues are now plentiful [See, e.g., (Lehnert et al. 1996; Golden et al. 1998; Adams et al. 2004; Batey et al. 2004; Golden et al. 2004; Serganov et al. 2004)]. Such structural conservation implies that the structure prediction problem is a meaningful one for functional RNA sequences — the three dimensional structures of these molecules are well-defined and indeed critical for understanding their biological function and evolution.

This chapter focuses on recent ideas for predicting the structure of an RNA sequence without experimental input. RNA secondary structures have been routinely ascertained prior to atomic-resolution experiments, often making use of phylogenetic covariation studies or easily obtained chemical footprinting profiles (Staehelin et al. 1968; Nussinov and Jacobson 1980; Zuker and Stiegler 1981). We therefore focus on the more difficult problem of modeling three-dimensional structures, especially regions involving non-canonical base-base and base-backbone interactions. Further questions such as the existence of alternative structures, the thermodynamics of these different states, the kinetics of self-assembly, and binding to proteins and other macromolecular partners are also important for understanding the biological behavior of RNA. While some current modeling approaches provide partial answers to these questions (Bowman et al. 2008; Ding et al. 2008), few rigorous experimental comparisons of simulations and non-structural experimental data (e.g., folding rates) have been reported. As with the much longer-studied but still unsolved problem of protein folding, we feel that the RNA structure prediction problem – involving comparison of hundreds of predicted atomic-level RNA coordinates to high resolution experimental models – currently provides the most appropriate test of computational approaches.

**4.3 3D RNA Modeling inspired by protein structure prediction**

Following efforts by several labs to produce manual 3D modeling packages [see, e.g., (Mueller and Brimacombe 1997; Massire and Westhof 1998; Martinez et al. 2008)], several automated modeling algorithms have become available. The methods differ greatly in their search methods and also in the assumptions made to approximate the physics of RNA self-assembly. Each algorithm offers a partial solution to the RNA tertiary folding problem; within the proper domain of application, each method reproduces existing experimental structures for at least some small systems. In this section, we first focus on the fragment assembly approaches studied in our group.

Our approaches draw inspiration from the most successful strategies taken in "knowledge-based" protein structure modeling: they make full use of approximate sequence homology, known structural motifs, and PDB-derived base-pair contact distributions. Fragment Assembly of RNA (FARNA) directly applies the Rosetta approach for de novo protein modeling (Das and Baker 2007) to RNA, a Monte Carlo conformational search making use of trinucleotide fragments drawn from a ~3000-nucleotide crystal structure of the large ribosomal subunit (Ban et al. 2000).

The assembly is guided by a coarse-grained scoring function, with parameters ascertained from the same ribosome crystal structure. The choice of using a knowledge-based potential was motivated by two considerations, both based on past experience with 3D protein modeling. First, we expected that deriving such a term from the database would ensure inclusion of physical terms that might be incorrectly modeled in a bottom-up, 'physics-based' derivation of the potential. For example, high-level effects of base aromaticity on hydrogen bond strength and the influence of the hydrophobic effect remain difficult to compute and calibrate, as they are for proteins (Simons et al. 1997).

The base-base interaction potential dominates the FARNA scoring function. We constructed this potential for each base interacting with the others, inspired by previous studies on classifying these interactions (Leontis and Westhof 2001; Sykes and Levitt 2005). After fixing one nucleobase at the origin, a total of six rigid body degrees of freedom describe the other base's orientation, three translational and three rotational. However, if this six-dimensional space is binned, the available statistics for base-pairing orientations in experimental structures is sparse; some bins have only one or two instances, and derived potentials can be noisy. A desire for a smooth landscape during the coarse Monte Carlo search led us to choose a two-dimensional reduction. [A similar choice was made in the Rosetta low-resolution potential for protein beta strand-pairings.] Thus, base pairing frequencies were tallied as a function of $x$ and $y$, i.e., the displacement of the centroid of the second base along directions parallel to the first base's plane (cf. Fig. 2a & Fig. 2b). Following a common (but not formally rigorous) recipe (Simons et al. 1997), a scoring function was derived by taking the log-ratio of the observed frequencies of these base-base orientations generated in de novo decoys compared to the frequencies seen in the ribosome structure. Separate terms favoring the appropriate base stagger ($z$) and colinearity of base normals were also implemented. Adding these terms one-by-one to further favor 'RNA-like' base-base arrangements led empirically to more accurate conformations of a test hairpin loop (Das and Baker 2007), at the expense of added computation to sample the more complex energy landscape. In fact, we also tested a higher dimensional representation including base-base rotation information that we expected to give better accuracy (parameterized on $x$ and $y$, as before, but also the base-base 'twist' in the $x$-$y$ plane), but fragment assembly with thousands of Monte Carlo cycles was unable to efficiently sample even simple hairpin loop conformations in this more complex energy landscape.

Besides base pairing, a second critical term was a potential increasing with decreasing distance separating two atoms, preventing them from overlapping. The functional form matches that successfully used in protein low-resolution modeling. Explicitly, the form is proportional to $(d^2 - d_0^2)^2$ for distances below a cutoff $d_0$ (3–5 Å, parameterized from the distance of closest approach seen in the ribosome crystal structure). Two other terms had less effect: a compaction term, proportional to radius-of-gyration, favors the well-packed conformations characteristic of experimentally-observed RNA structures, but such conformations are already well-favored by the base-base interaction potential. Finally, a base-stacking term rewards conformations that stack bases in a parallel orientation; here, the stacking geometries already ensured by constructing models from ribosome fragments made the additional potential largely superfluous.

As should be apparent from the discussion above, derivation of a knowledge-based scoring function is a heuristic procedure, and the best test of such potentials is whether they result in more accurate de novo models. In favorable cases (under 20 residues), FARNA can sample and select out moderate resolution (2-4 Å all-atom root-mean-square-deviation, RMSD) models, as is illustrated for the GCAA hairpin loop in Fig. 1b (PDB: 1ZIH) (Jucker et al. 1996). Nevertheless, many contain steric clashes and poorly optimized hydrogen bonds. Furthermore, in larger systems, the scoring function fails to discriminate these <4 Å accuracy conformations from non-native decoys, although the accuracy can be improved by using experimental data (Das et al. 2008).

As with Rosetta approaches for protein structure prediction, the FARNA approach to RNA modeling is computationally expensive. The computational time to create a single model for a 12-nucleotide motif like the GCAA hairpin loop is approximately 10 seconds on an Intel Xeon 2.33 GHz processor; typical runs, however, involve the generation of at least 5,000 models, requiring 14 CPU-hours. The computational expense for generating single models of larger sequences scales approximately as the number of nucleotides.

The most rigorous test of FARNA has been the blind modeling of a 74 nucleotide RNA transcript containing three stems from a bacterial ribosome, for CAPRI (Critical Assessment of Prediction of Interactions) target T33, a complex of this RNA and a methyltransferase. Biochemical data suggested a large conformational difference between the structure of this RNA when bound to RlmAII compared to its known structure within the ribosome (Fig. 3a). We therefore applied automated *de novo* modeling to the RNA, with the hopes of selecting an accurate conformation through post facto docking to the protein component.

Although the modeling did not converge at high resolution (< 2 Å RMSD), low energy configurations shared an overall global fold that was quite distinct from the ribosome-bound fold, especially in the helix-helix geometry at the molecule's three-way junction (Fig. 3b). Subsequent release of the protein-bound RNA crystallographic model (Fig. 3c) revealed that a conformational rearrangement indeed occurs. The blind prediction was accurate at modest resolution, 5.4 Å RMSD over C4´ atoms (residues 694-702, 730-737, and 759-767), compared to 12.4 Å in the previously available ribosome-bound conformation (Fleishman 2010). [The unavailability of the crystallographic coordinates to the public at the time of writing preclude presentation of the protein-bound model in this chapter.]

**4.4 A wealth of 3D RNA modeling approaches**

There are now several algorithms for *de novo* modeling of RNA structure in addition to the fragment assembly approach described in the previous section, spanning a spectrum from more knowledge-based methods to more physics-based methods. Before discussing limitations of our fragment assembly approach, we survey these alternative methods, comparing results on one widely modeled sequence, the GCAA tetraloop, and, in the next section, the sarcin-ricin loop.

Like FARNA, the accuracy of the MC-Fold/MC-Sym pipeline (Parisien and Major 2008) depends on the available set of experimentally solved RNA structures. MC-Fold uses small RNA building blocks (nucleotide cyclic motifs, NCM) that are pieced into a two-dimensional representation of the RNA. The result is essentially an extended secondary structure (2D-3D) that includes both canonical and non-canonical non-Watson-Crick base pairs; it is the optimum of a Bayesian scoring function derived from the previously tallied frequencies of NCMs in experimental structures. This two-dimensional model is then submitted to MC-Sym (Major et al. 1991), a pioneering modeling method that generates three-dimensional structures consistent with the inputted secondary structure, often with outstanding accuracy (better than 2 Å all-atom RMSD; Fig. 1d). Like FARNA, the MC-Fold/MC-Sym method does not require prior determination of secondary structure or experimental constraints but can accept and greatly benefit from these additional data (McGraw et al. 2009).

While both FARNA and MC-Fold/MC-Sym reward previously seen base-pairing geometries, several algorithms are less reliant on the existing databases. For instance, Discrete Molecular Dynamics (DMD) (Ding et al. 2008) uses an efficient molecular dynamics engine to sample coarse-grained RNA structures. In numerous cases, DMD achieves native-like structures (~4 Å C4´ RMSD, with some models as low as 1.9 Å ; Fig. 1e) without explicit calibration on any RNA conformations aside from canonical helices; energetic parameters are calibrated to classic thermodynamic experiments on RNA helix formation (Xia et al.

1998). Similarly, NAST (Jonikas et al. 2009) uses coarse-grained molecular dynamics with a force-field parameterized to reproduce canonical helices. Both of these approaches use molecular dynamics strategies that, by construction, do not explicitly model noncanonical regions. The accuracy of these methods can be improved through the use of experimental constraints (Gherghe et al. 2009; Jonikas et al. 2009).

At the other end of the spectrum, all-atom molecular dynamics approaches do not make use of information from structural databases, aside from corrections to the underlying energy function to stabilize experimental conformations (Foloppe and MacKerell Jr 2000). For example, all-atom molecular dynamics simulations have been used (Sorin et al. 2002; Bowman et al. 2008; Garcia and Paschek 2008) to investigate small RNA hairpin loops. Experimental hairpin loop structures appear to be stable in several solvation models (Sorin et al. 2002; Bowman et al. 2008), and the native-like secondary structure appears to be reachable from randomized conformations. Nevertheless, the precise details of non-canonical loop geometry may not be recapitulated by presently available force fields (see, e.g., (Ditzler et al. 2010)).

**4.5 Case Study: Sarcin-Ricin Loop Suggests Limitations of Current Methods**

The preceding survey of methods suggests that residue-level, and occasionally atomic-level, accuracy can be achieved in three-dimensional RNA modeling by a multitude of approaches. Yet, the RNA structure prediction problem is far from solved. The computational methods described so far cannot reliably produce high-quality models of an arbitrary RNA, a point we demonstrate with a long-studied model system. The structure of the sarcin-ricin loop, revealed in exquisite detail by X-ray crystallography (Fig. 4a and Fig. 5, PDB: 1Q9A), contains a tightly intermeshed array of hydrogen bonds. Within the seven nucleotides that form the core of this motif, there are 11 hydrogen bonds present (6 base-base, 4 base-phosphate, and 1 base-sugar), resulting in an average of 1.57 hydrogen-bonds per nucleotide, greater even than the 1.50 hydrogen-bonds per nucleotide in a repeating GC helix. The structural stability of this small motif has made it a paradigmatic system for experimental studies (Endo et al. 1991; Seggerson and Moore 1998) and a test case for evaluating current computational algorithms.

Applying FARNA to the sarcin-ricin loop yields mixed results. While FARNA produces native-like models (under 2 Å RMSD), the knowledge-based scoring function fails to distinguish these models from incorrect models (Fig. 4b and Fig. 4c). The same motif proves a challenge for other algorithms as well. Each of the top twenty models produced by MC-Fold has an incorrect base pair, suggesting limitations in the knowledge-based scoring function (Fig. 4d). As was the case with FARNA, several of the poorer scoring MC-Fold models (Fig. 4e) contain the correct base pairs and topology. Interestingly, using slightly different homologous sequences leads to better performance with both FARNA (unpublished data, PS, RD) and MC-Fold/MC-Sym (Parisien and Major 2008).

DMD, followed by all-atom reconstruction (Sharma et al. 2008), likewise cannot reproduce this structure at high resolution (Fig. 4f). These three algorithms use very different modeling strategies for RNA conformational sampling—a smoothed energy landscape (FARNA), a two-dimensional NCM description (MC-Fold), and a coarse-grained representation (DMD). Yet, all three algorithms fail on the same model system. A final class of algorithms, all-atom molecular dynamics simulations, satisfy a basic consistency check: the sarcin-ricin loop structure is stable in several tested force fields over the nanosecond time scale (Spackova and Sponer 2006). However, such simulations have not yet been carried out on the long timescales necessary for folding and discriminating complex RNA structures *de novo* (Pérez et al. 2007). Thus, it is presently unclear whether (and at what resolution) molecular dynamics can recapitulate larger experimental structures in simulations started from random conformations; indeed, a few cautionary tales have suggested that noncanonical motifs are unstable in existing MD force fields (Fadrná et al. 2009).

**4.6 What are the bottlenecks?**

The situation in RNA structure prediction bears some similarities to the state of protein modeling. Several algorithms are able to reproduce a handful of known small structures at reasonable resolution. Nevertheless, foundational bottlenecks prevent the prediction of known complex structures, as described above, despite the diversity of approaches being applied. The failure on known structures lowers our confidence that the existing approaches can be used to accurately predict new structures. Here, we describe three hypotheses for bottlenecks that need to be overcome.

*4.6.1 Computational Sampling*

Despite their differences, all RNA modeling algorithms proposed to date share a major difficulty in computational sampling, especially if they seek atomic resolution. For example, the trial runs on the sarcin/ricin loop above were made possible by the small size (<30 residues) of the tested motif; modeling of larger segments of the ribosome, much less the entire large ribosomal subunit, is presently difficult. The root of this problem was first discussed more than 40 years ago, when Levinthal noted that the conformational space available to a biomolecule is astronomical ($10^{100}$) and grows exponentially with the number of residues (Levinthal 1968). Forty years later, algorithms – for protein as well as RNA structure modeling – continue to face Levinthal's Paradox, as they typically involve a near-random search through conformation space.

As noted above, the difficulty of conformational sampling has prevented all-atom molecular dynamics approaches from demonstrating *de novo* recapitulation of RNA structure at high resolution. Other approaches are less expensive, but still require high performance computing. For example, FARNA calculations, even though constrained by experimental data, required approximately 10,000 CPU-hours on the Rosetta@Home distributed computing project to model the P4-P6 domain of the *Tetrahymena* ribozyme (160 residues) to ~13 Å accuracy (Das and Baker 2007). The barrier of ~100 nucleotides is particularly worrisome because several of the most biologically and medically important RNAs – including viral genomes (Watts et al. 2009) and untranslated regions of mRNA transcripts [see, e.g., (Penny et al. 1996; Birney et al. 2007)] – can exceed thousands of residues in length. Several methods (MC-Sym, DMD, NAST) appear significantly less expensive than FARNA; nevertheless, each algorithm is expected to encounter a conformational sampling bottleneck for some length of RNA. Detailed presentations of these length limits are not yet available but would certainly be valuable for users of the algorithms.

*4.6.2 Over-reliance on Existing Structures*

One common approach to ameliorate the computational sampling bottleneck is to restrict the search to torsion angles of base pairing combinations drawn from the experimental database of known RNA structures. On one hand, the success with the tetraloop motif can perhaps be attributed to not just its simplicity but its overall frequency in the database of experimental RNA structures. On the other hand, we might expect poor performance on novel sequence motifs if they exhibit torsional geometries that are at low frequency or are absent in current databases.

As an extreme illustration, the FARNA method fails to recover high resolution models for motifs such as the kissing loop from the purine-binding riboswitch, unless this structure or its homologues are included as sources of fragments (Fig. 6). The intricate base-pairing and base-stacking network formed by the two loops requires the individual nucleotides to adopt highly specific conformations. Consistent with this observation, the backbone conformations of three nucleotides in this motif are not on the list of commonly observed backbone rotamers compiled by the RNA Ontology Consortium (Richardson et al. 2008). The lack of native-like fragments in the fragment library prevents FARNA from generating models that are within 5 Å RMSD of the crystal structure.

Beyond adversely limiting the conformational space, reliance on existing structures can also cause problems in ranking models by available scoring functions. A widely studied RNA motif involves a quadruplex of G nucleotides forming parallel base pairs [see, e.g., (Mashima et al. 2009)], yet available ribosome crystal structures contain no such guanosine arrangements. This leads to a known deficiency in FARNA (cf. Fig. 2c & 2d); namely, some known G-G base interactions are not rewarded by the FARNA scoring function, and quadruplexes cannot be modeled. The scoring function can be reparameterized with the entire non-redundant crystallographic RNA data set (rather than just a single ribosomal structure), but will continue to miss important interactions, such as those involving protonated C's or A's, which are important for stabilizing RNA motifs but that are rare in the entire set of RNA structures. Ironically, while the ultimate goal of structure prediction is to model motifs that have not yet been observed experimentally, it is these novel structures that knowledge-based algorithms have the most difficulty predicting.

*4.6.3 Simplified Representation*

Perhaps the central shared bottleneck of the various *de novo* approaches discussed so far is the use of a simplified representation. Searching RNA conformations in all-atom detail requires attention to hydrogen bonds and packing interactions at the Angstrom level; algorithms to directly and efficiently sample conformations at this level of detail are not available. Instead, as is the case in protein structure modeling, *de novo* RNA modeling algorithms typically resort to a coarse-grained representation to carry out large-scale conformational search. In FARNA, the energy function is highly smoothed; MC-Fold uses a two-dimensional secondary-structure-like representation; and NAST and DMD both use reduced-atom models to accelerate molecular dynamics sampling.

These methods inevitably neglect certain details of RNA structure – but in the case of the sarcin-ricin bulged-G motif, these details cannot safely be ignored. Because the FARNA scoring function represents base-pairing and base-stacking interactions at the base-centroid level, the atomic details of individual hydrogen bonds are not represented—leading to an inability to select the native conformation of this motif (Fig. 4c).

The discrimination of realistic RNA conformations should be possible using all-atom physical potentials, and indeed such potentials have been the key feature in recent blind *de novo* Rosetta predictions of protein structure at near-atomic accuracy (Rohl et al. 2004). For RNA, the backbone torsional combinations seen in real structures are those physically allowed by sterics and torsional constraints; further, base pairing patterns follow the "laws" of hydrogen-bonding as well (Leontis and Westhof 2001). [See also Figs. 2b & 2d, which were generated by exhaustively sampling base-base arrangements, scored with the Rosetta full-atom potential.] Recognizing the power of all-atom potentials, several groups have explored the refinement of automatically generated pools of low-resolution structures with all-atom potentials (Sharma et al. 2008; Jonikas et al. 2009).

In our own recent work (Das et al. 2010) we have found that the full-atom Rosetta RNA force-field can correctly refine and discriminate near-native structures for more than a dozen non-canonical motifs, including the bulged-G region of the sarcin-ricin loop structure (Fig. 7). The Rosetta RNA force-field is essentially the same as used in protein structure prediction, with physics-based van der Waals (Rohl et al. 2004) and hydrogen bonding terms (Kortemme et al. 2003) supplemented with a torsional potential inferred from the ribosome $\alpha$, $\beta$, $\gamma$, $\delta$, $\epsilon$, $\zeta$, and $\chi$ torsion angles; a desolvation penalty for polar groups that models how neighbor atoms occlude water; and a weak carbon-hydrogen bonding term. [The latter two terms appear to improve protein structure prediction as well.] The overall structure modeling procedure (Fragment Assembly of RNA with Full-Atom Refinement, FARFAR) doubles the time of the previous FARNA method, to 21 seconds on an Intel Xeon 2.33 GHz processor for the 12-residue GCAA hairpin loop. Further independent tests of the approach, involving the "re-design" of RNA sequences that stabilize known backbone conformations, gave higher native sequence recoveries than low resolution potentials (Das et al. 2010). Most importantly, the calculations gave blind predictions for thermostabilizing non-canonical mutations that were validated in subsequent experiments (Fig. 8). We hope that the free availability of these algorithms to academic users (as part of the Rosetta software suite) will encourage their testing and development in the broader community.

The demonstrations of atomic accuracy structural modeling and design are exciting steps, but also confirm that conceptual advances in conformational sampling are much needed. In the published benchmark (Das et al. 2010), the structures of half of the 32 noncanonical motifs could be recovered at atomic accuracy. For most cases in the other half, sampled models produced scores worse than the experimental structure, indicating that conformational sampling was not efficient. In particular, motifs beyond approximately 12 residues in length are still difficult to sample at the Angstrom-level resolution required for high accuracy discrimination; a similar conformational sampling issue remains the major bottleneck in *de novo* prediction of protein structure and docking (Das et al. 2009; Kim et al. 2009; Raman et al. 2009). We also expect there to be missing physics in the all-atom Rosetta energy function, due to the neglect of explicit metal ions and water, of terms to modulate the strength of base stacking, of long-range electrostatic effects, and of conformational entropy. However, more effective conformational search procedures will be needed to establish whether these effects are critical for discriminating high-accuracy models from non-native models. Based on the three major issues discussed above, we are currently focusing on approaches that enumeratively search realistic conformations of biomolecules, are independent of previously solved structures, and bypass coarse-grained search stages.

**4.7 Future Directions/Community Wide RNA experiments**

Given the promising algorithms currently under development, it is reasonable to expect improved *de novo* methods for (small) RNA structure prediction in the next few years. However, once such novel algorithms are developed, they must be rigorously tested before they will be accepted and used by the wider RNA community.

In particular, the present cycle of algorithm development, testing, and publication inevitably pressures scientists to present their results in an overly optimistic and sometimes uncritical fashion. A blind, CASP-style competition to systematically assess the performance of RNA 3D structure prediction algorithms will therefore be crucial for future progress. We request the cooperation of experimentalists to make available, prior to publication of an experimental atomic-resolution RNA structure, the nucleotide sequence of the solved molecule and to provide a deadline for modelers to submit solutions. We can expect that objective evaluation of such trials will encourage thoughtful and open discussion of the strengths and limitations of current approaches and engender new collaborations between modelers and experimentalists.

For a CASP-style experiment to be interesting and useful, truly novel targets must be included. We note that at least three classes of such targets are already available to the modeling community. First, the growing interest in functional RNAs has led to crystallographic analyses of several new, large riboswitches. Although the sizes of these RNAs (>100 residues) puts them out of the reach of current algorithms, sub-motifs (such as internal loops and junctions) may fold in a manner largely independent of ligand binding or other interactions. The L2-L3 motif from the adenine riboswitch is such an example and is stable independent of adenine binding (Mandal and Breaker 2004; Serganov et al. 2004); internal multi-helix junctions from the glycine riboswitches may provide another presently unsolved set of tests (RD, unpublished data). Biochemical identification of these sub-puzzles, combined with the rapid rate at which these functional molecules are being crystallized, suggest that they are excellent targets for blind prediction.

Using *in vitro* evolution to redesign existing motifs provides another class of novel targets (Fig. 9a). A compelling example comes from the determination of optimal receptor motifs that specifically bind GNRA tetraloops (Costa and Michel 1997; Geary et al. 2007). These new motifs are less than a dozen residues in size, and most have presently unknown structure, making them ideal targets for current modeling approaches. Furthermore, these targets offer the prospect of rapid experimental validation. Given the growing number of ribozyme and riboswitch structures with the classic 11-nucleotide receptor motif for GAAA, the experimental structures of the alternate tetraloop-receptors may be attained by their substitution into RNA scaffolds that are known to be crystallizable (Pley et al. 1994; Cate et al. 1996; Ye et al. 2008).

Finally, there is a large body of work focused on sequences that bind small and large molecules, again isolated through in vitro evolution. Many of these functional sequences are small – only 13 nucleotides in the case of an L-tryptophan aptamer (Majerfeld and Yarus 2005) (Fig. 9b)—again bringing them closer to the reach of all-atom computational modeling. Furthermore, their small size should permit their rapid experimental characterization by modern NMR approaches, ensuring a nearly unending supply of targets for blind trials.

**4.7 Conclusions**

Predicting the structure of an arbitrary RNA sequence remains an unsolved problem. A number of algorithms can rightly claim success in specific cases, including some blind tests; but a general solution has yet to appear, even for small sequences. Present methods are limited by computational sampling, over-reliance on previously solved experimental structures, and the use of coarse-grained or reduced representations. Recent progress, especially in all-atom refinement and design, makes us particularly excited about the future; a solution to RNA structure prediction appears more and more feasible. We propose that the time is ripe for the creation of a community-wide CASP-style experiment, where groups compete to produce blind models of RNAs about to be solved by crystallography or NMR. The prospect of such blind trials bodes well for the maturing and eventual practical impact of the RNA structure prediction field.

**Figure 1.** Models of the GCAA tetraloop structure. Secondary structure annotations follow the convention of Leontis and Westhof (Leontis and Westhof 2001) and were prepared with the aid of RNAmlView (Yang et al. 2003) and FR3D (Sarver et al. 2008). (a) NMR structure (PDB: 1zih). (b) Lowest energy model of 5000 fragment assembly models [2.1 Å]. (c) The knowledge-based FARNA scoring function correctly ranks a near-native model as the lowest energy state (red circle). (d) Lowest RMSD of 3 MC-Sym models [1.7 Å]. The lowest energy secondary structure as determined by MC-Fold is used as MC-Sym's input. (e) Lowest RMSD of 20 DMD models [1.9 Å]. Reported RMSDs are calculated over all heavy atoms with respect to the first member of the NMR ensemble.

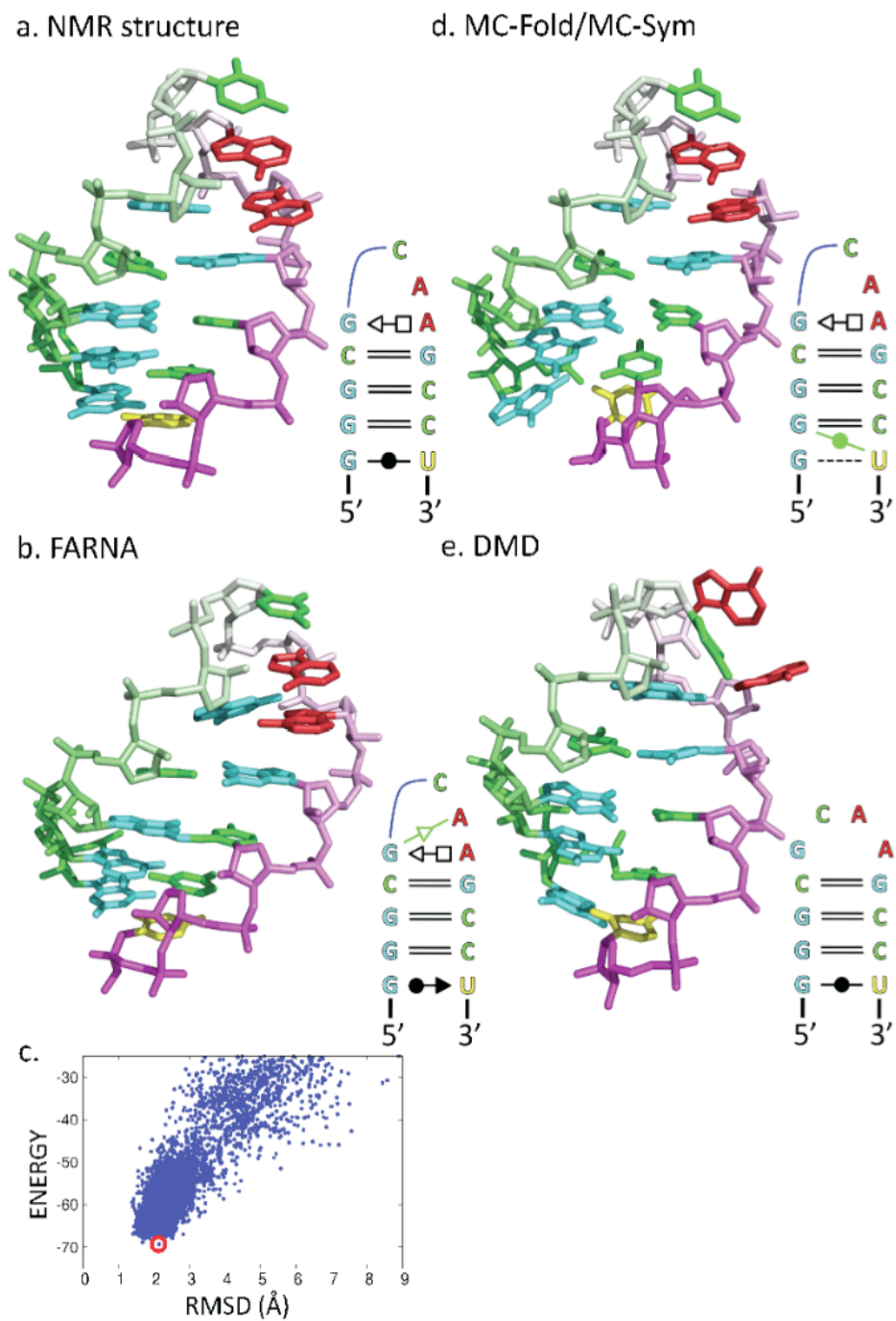

**Figure 2.** Comparisons of the knowledge-based force field used in Fragment Assembly of RNA (FARNA) to base-base orientations generated by *de novo* sampling. The distribution of uracil bases around adenosine (filtered for configurations in which the base normals are antiparallel), based on (a) the crystal structure of the large ribosomal subunit (1JJ2), as used in the FARNA scoring function (Das and Baker 2007); and (b) a calculation enumerating all physically reasonable base-base orientations, scored with the high-resolution Rosetta force field. The three common antiparallel A-U configurations are seen with both approaches. In contrast, the distribution of parallel guanosine-guanosine base pairs as inferred from the ribosome (c) does not recapitulate all physically allowed configurations (d).

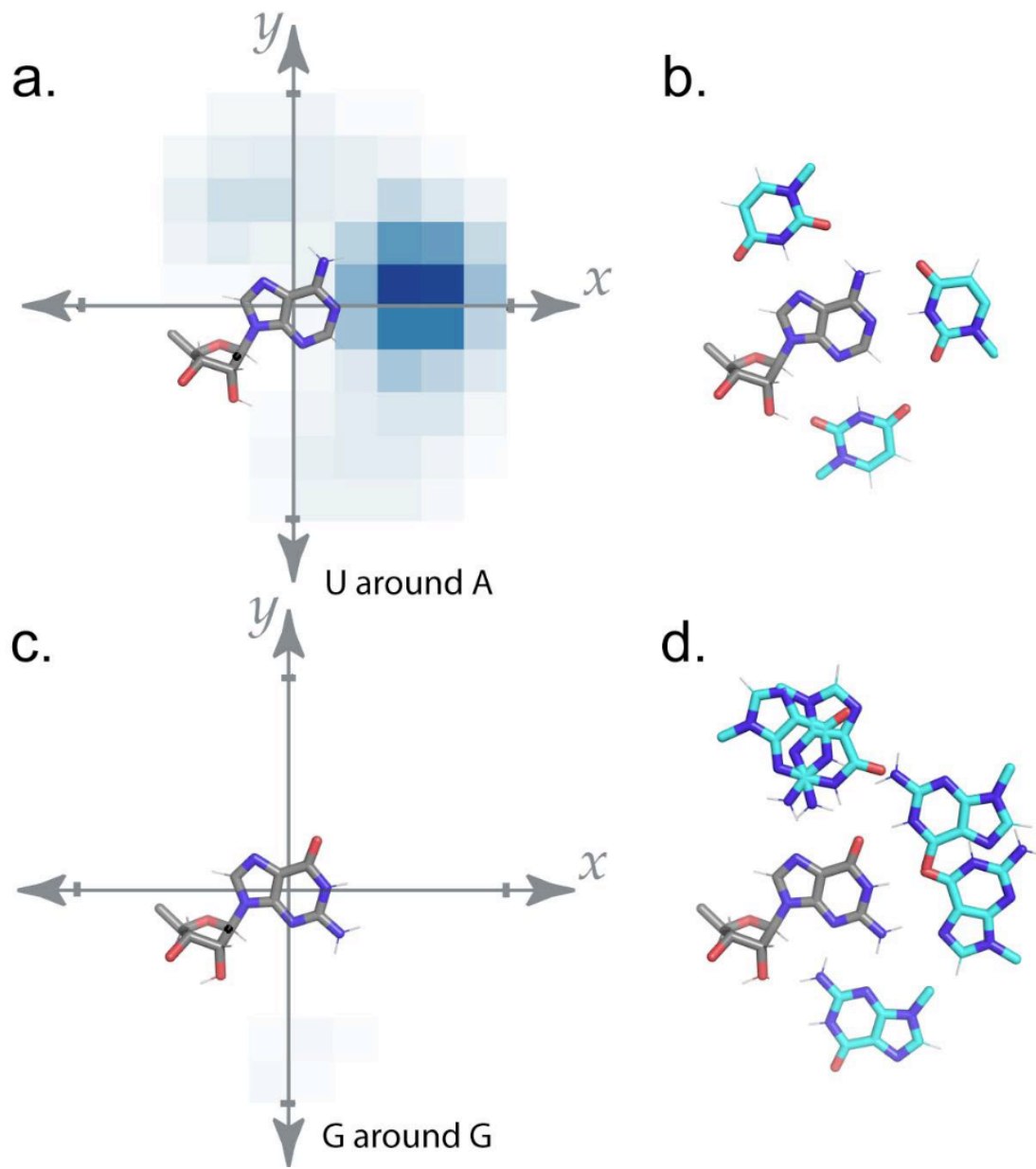

**Figure 3.** De novo modeling of target T33 in the Critical Assessment of PRotein Interactions (CAPRI) trials, the complex of an rRNA segment and a methyltransferase (Fleishman 2010) (A) The previously available structure of the three-helix junction in the context of the *E. coli* ribosome. (B) Representative *de novo* model generated by Fragment Assembly of RNA (FARNA) suggested a large conformational change, with additional support from full-atom refinement as well as low-resolution docking simulations with the protein target (not shown). The subsequently released crystallographic model of the RlmAII-bound RNA confirmed the conformational rearrangement but cannot be presented here because the coordinates are not yet publicly available.

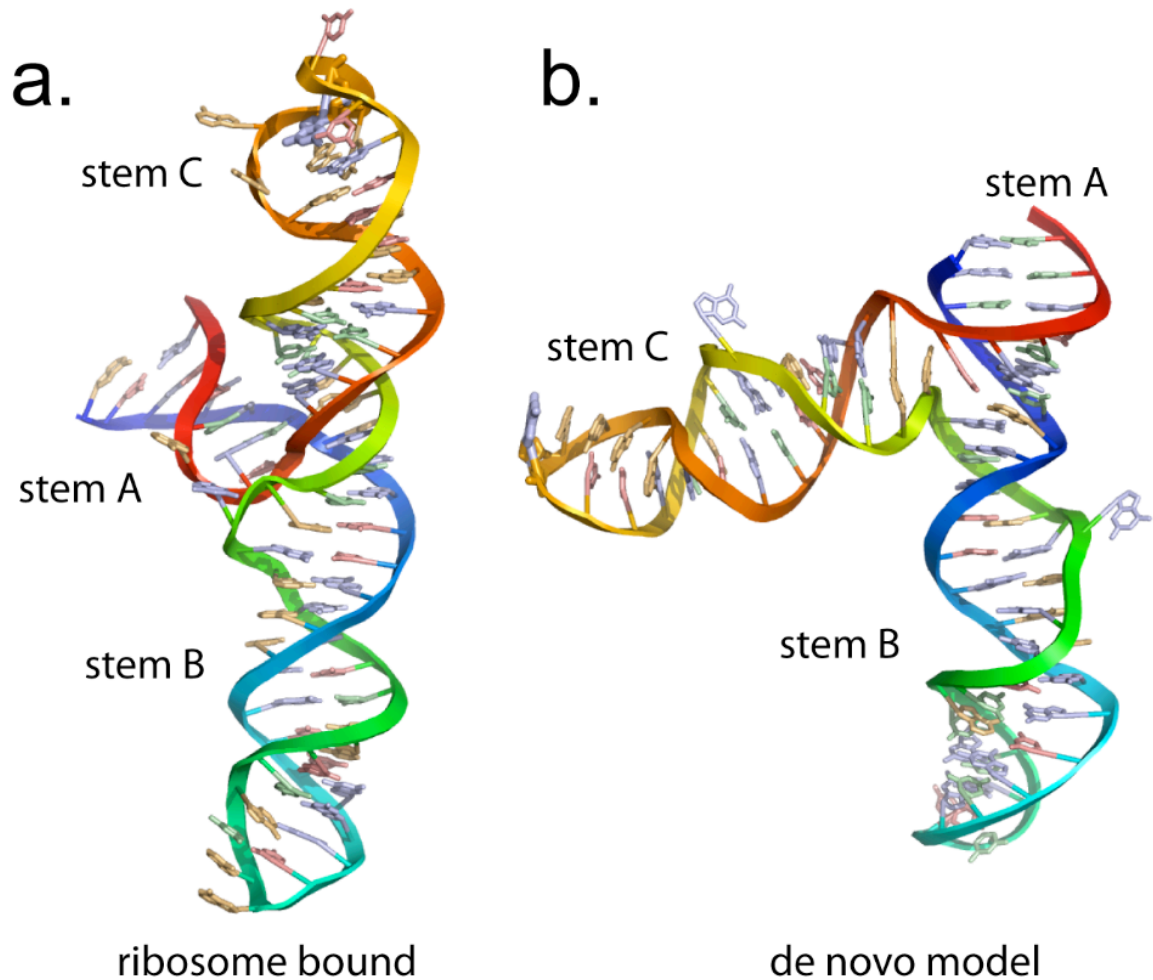

**Figure 4.** Models of the sarcin-ricin loop structure. The depicted three-dimensional structures focus on the bulged-G motif region (red box in secondary structure) where all three algorithms fail. Interestingly, all three algorithms predict that G9 and C20 form a Watson-Crick base pair (red circle) which is absent in the crystal structure (a) Crystal structure (PDB: 1Q9A). (b) Lowest energy model of 50,000 fragment assembly models [6.198 Å]. (c) The knowledge-based FARNA scoring function is too coarse-grained and incorrectly ranks the non-native model (red) as better in score compared to near-native model (green). (d) Lowest RMSD of 100 MC-Sym models [3.809 Å]. (e) The lowest energy secondary structure as determined by MC-Fold is used as MC-Sym's input. This secondary structure contains incorrect base-pairings, but the native (correct) secondary structure is ranked #24 (red) by MC-Fold. (f) Lowest RMSD of 20 DMD models [4.704 Å]. Reported RMSDs are calculated over all heavy atoms with respect to the crystal structure.

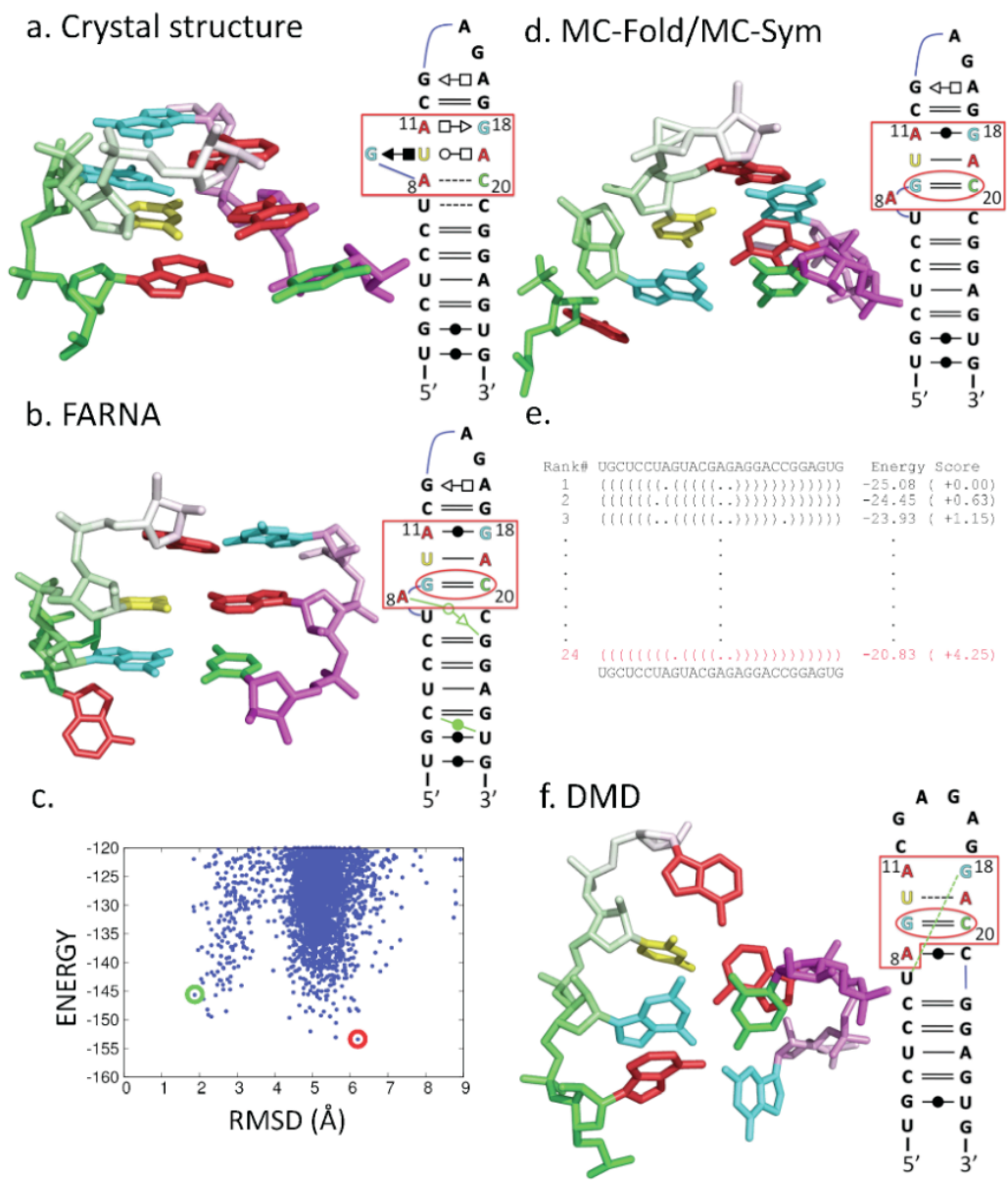

**Figure 5.** Hydrogen bonding network at the bulged-G motif in the the crystal structure of the sarcin-ricin loop (PDB: 1Q9A). Base-phosphate hydrogen bonds in the bulged-G motif region are annotated following a recently proposed convention (Zirbel et al. 2009). (a) Upper section of the motif. (b) Lower section of the motif. To display all the hydrogen bonds, the views in (a) and (b) are rotated by 180° with respect to each other. The experimentally observed hydrogen bonds are shown as red dashed lines. Within the seven residues which form the core of this motif (A8-G9-U10-A11/G18-A19-C20), there are 11 unique hydrogen bonds (6 base-base, 4 base-phosphate and 1 base-sugar), averaging to 1.57 hydrogen bonds per nucleotide. This is slightly greater than the number of hydrogen bonds in a repeating GC helix (1.50 hydrogen bonds per nucleotide). In contrast none of the models generated by DMD, FARNA or MC-FOLD shown in Figure 4 have greater than 1.0 hydrogen bonds per nucleotide in this same region. Furthermore, in the models generated by these three algorithms, very few of the hydrogen bonds are of the base-phosphate or base-sugar type.

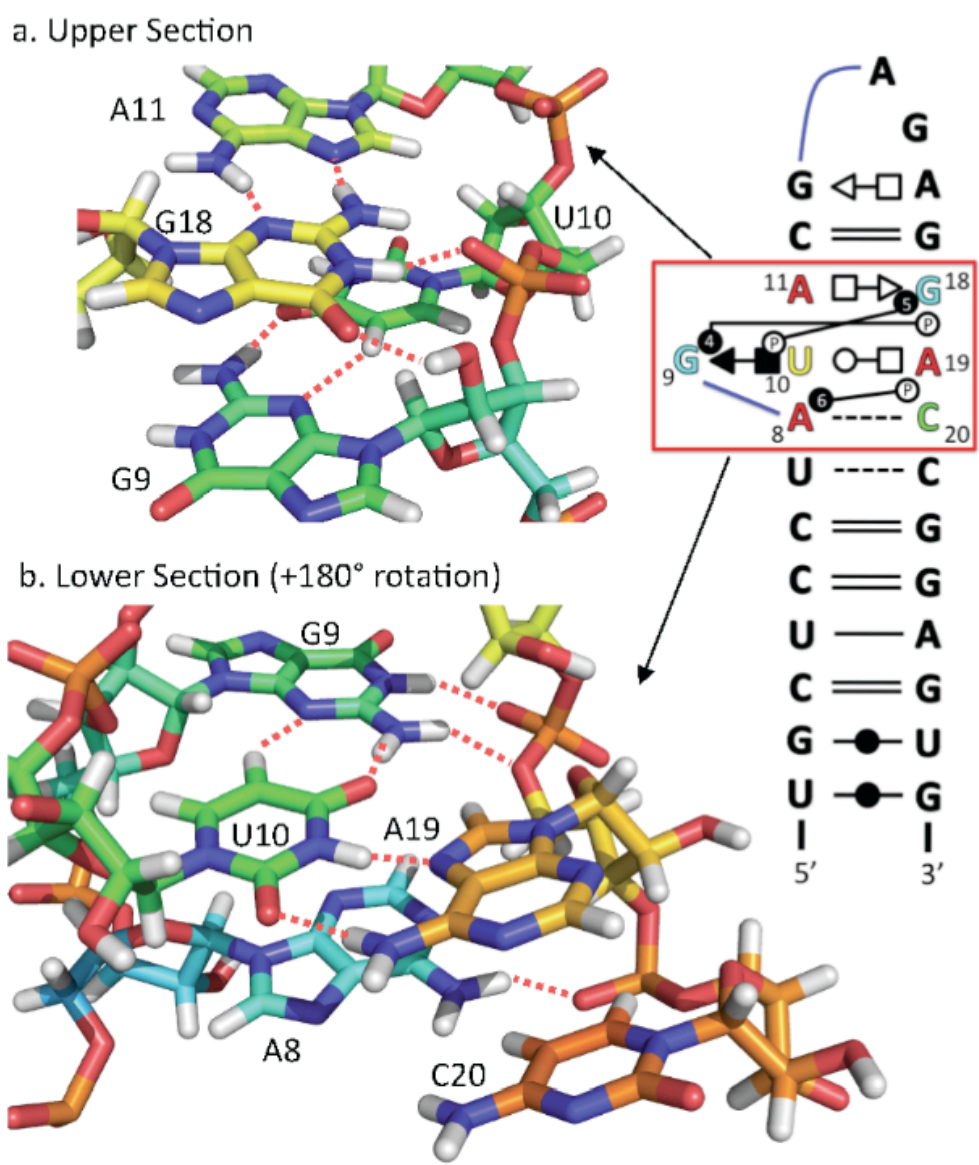

**Figure 6.** The L2-L3 tertiary interaction from the purine riboswitch is poorly sampled using FARNA. (a) The crystal structure (PDB: 2EEW). (b) The lowest energy model of 5,000 FARNA models. (c) Secondary structure annotation of the crystal structure. (d) An energy score vs. RMSD plot shows that FARNA fails to generate models that are closer than 5 Å to the crystal structure. Examination of the large ribosomal subunit fragment library reveals the lack of near-native fragments at many residue positions. This observation suggests that the limitations in this case are a combination of both computational sampling and the knowledge-based limitations of an incomplete fragment library.

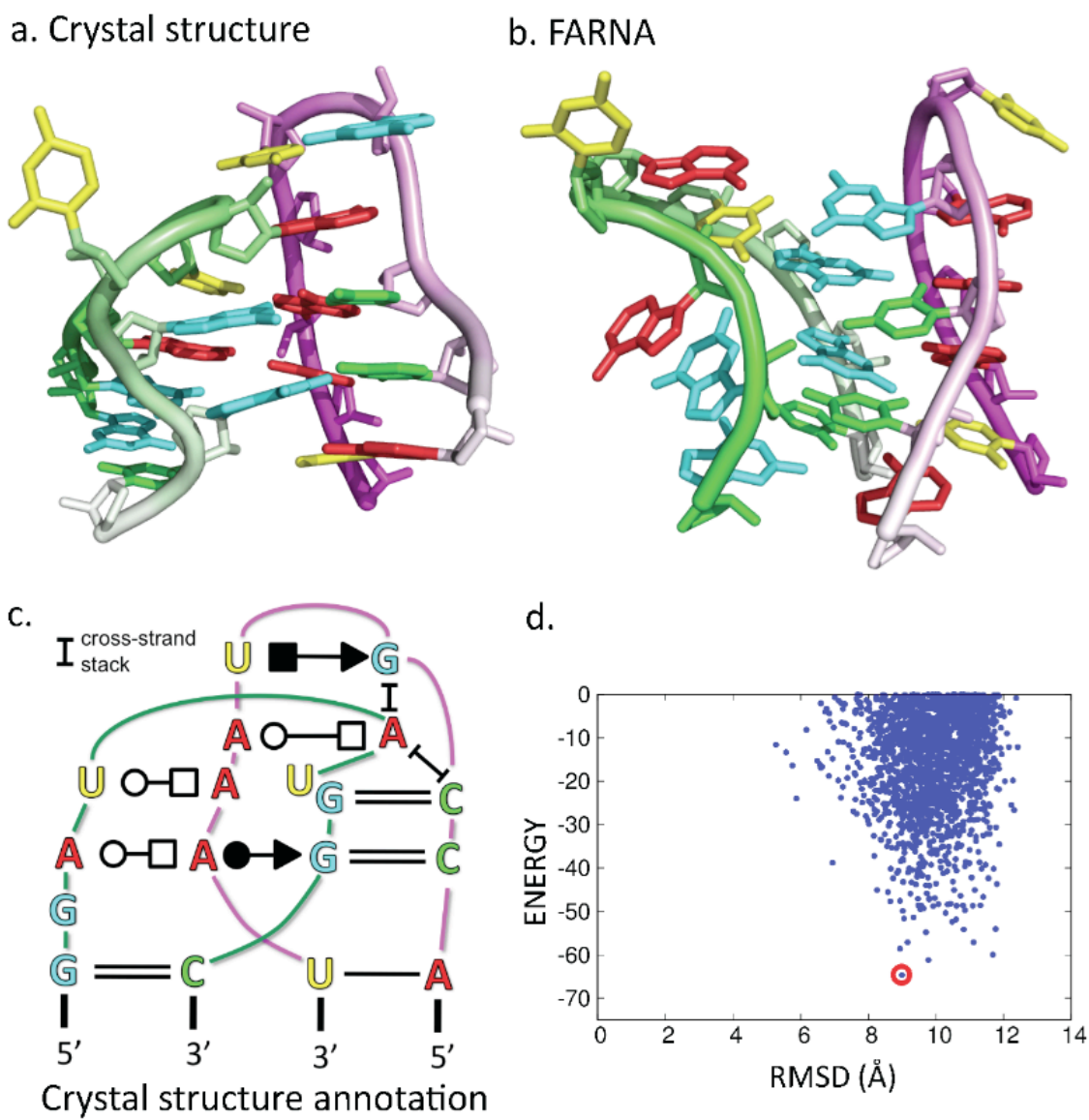

**Figure 7.** Successful modeling of the sarcin-ricin loop with FARFAR. Previous attempts to model the sarcin-ricin loop with FARNA indicated that while fragment assembly generated several models with native-like base-pairing geometries, the knowledge-based FARNA energy scoring function was too coarse-grained and incorrectly ranks the non-native model as better in score compared to near-native model (see Fig. 4). (a) In contrast, the Rosetta full-atom force-field used in Fragment Assembly of RNA with Full-Atom Refinement (FARFAR) more accurately models the energetics of the hydrogen bonding network in the bulged-G motif. When the same 50,000 fragment assembly models are refined (minimized) and scored in the full-atom force-field, the near-native model (green) is correctly ranked as the lowest energy state. (c, d) This near-native model has a 1.798 Å RMSD with respect to the whole sarcin-ricin crystal structure and even a lower local RMSD of 1.038 Å when aligned just over the bulged-G motif nucleotides.

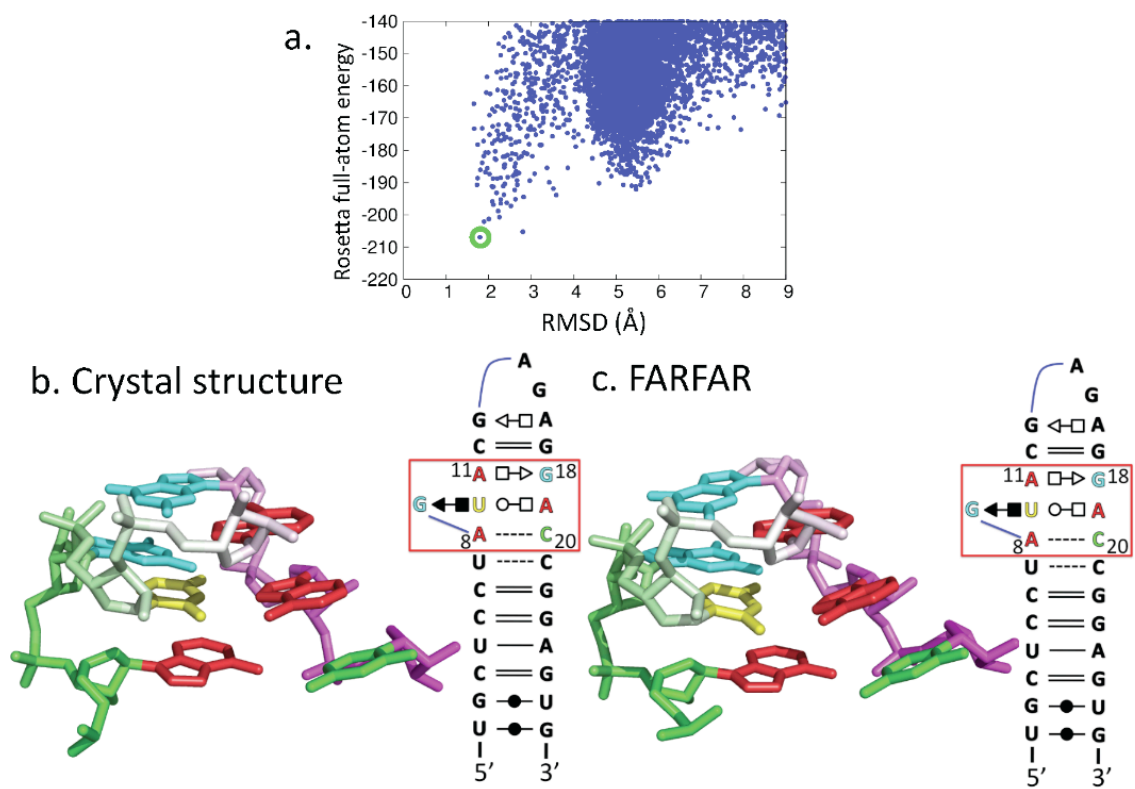

**Figure 8.** Automated 'redesign' of an RNA noncanonical motif. (a) The crystal structure of the most conserved domain of the signal recognition particle RNA (PDB: 1LNT). Sidechains from this structure were completely stripped away and then rebuilt, sampling all possible sequences, guided by the Rosetta full-atom force field. (b) Two mutations (highlighted with arrows) were discovered in the library of designs that occurred more frequently than in an alignment of natural sequences from all three kingdoms of life. (c) The corresponding secondary structure of the wild-type and the mutant. (d) Experimental structure mapping measurements verify the stabilization of the motif by the two mutations (less $Mg^{2+}$ required for folding) (Das et al. 2010).

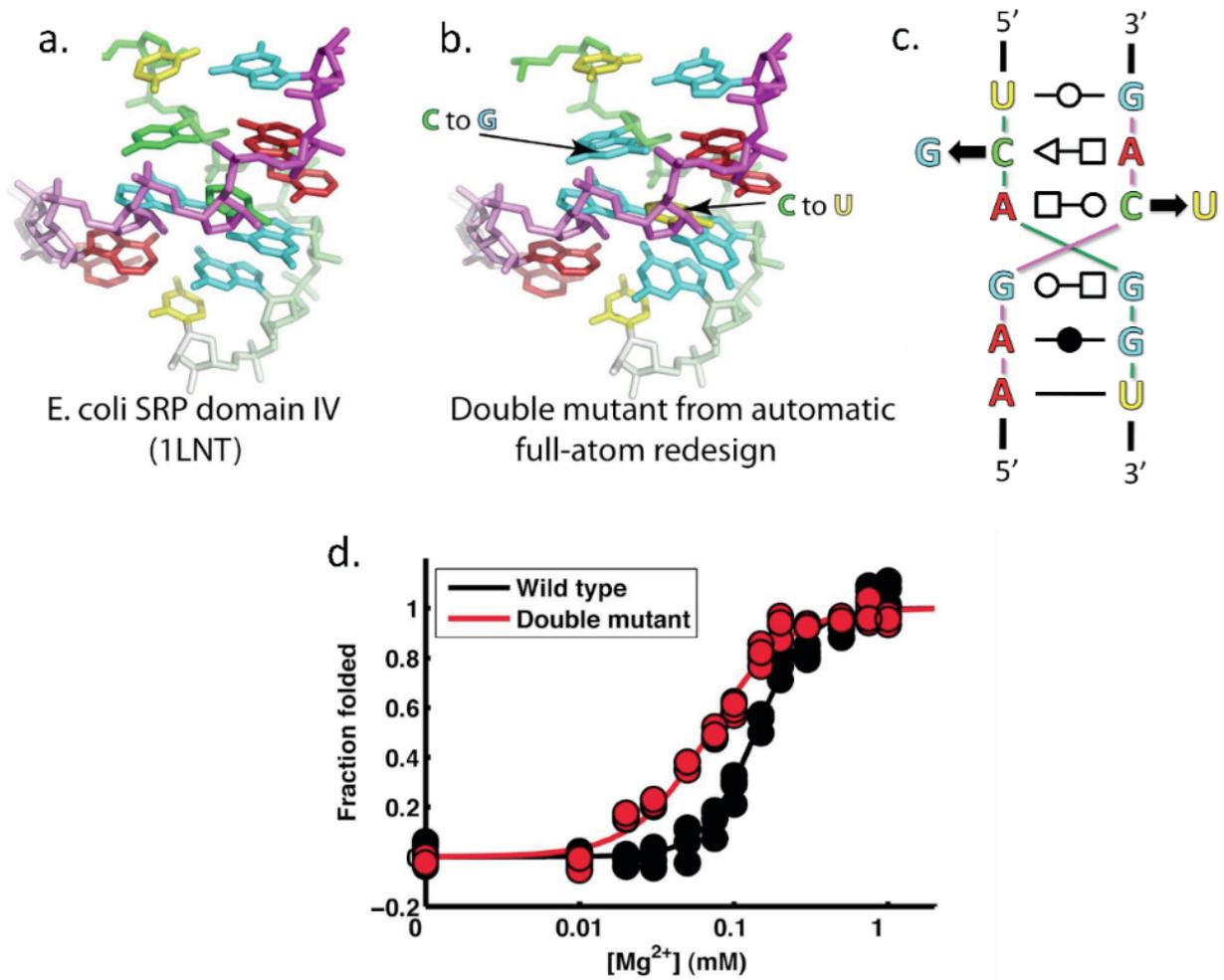

**Figure 9.** Examples of novel RNA motifs with unknown structure. These motifs' small size, lack of homology to known structures, and potential ease of experimental validation make them ideal tests for computational algorithms. (a) The naturally occurring GAAA/11-nucleotide tetraloop receptor with known experimental structure (Pley et al. 1994; Cate et al. 1996; Ye et al. 2008). (b) *In vitro* selected tetraloop receptor motifs (Costa and Michel 1997). The binding affinities and specificities of these tetraloop receptors are markedly different from those of known naturally occurring tetraloop receptors. It is postulated that these novel binding specificity patterns are due to some (as yet undetermined) interactions between the second base of the tetraloop (red) and the nucleotides in the asymmetric loop of the receptor (green). (c) Binding site of an *in vitro* selected L-tryptophan binding aptamer (Majerfeld and Yarus 2005).

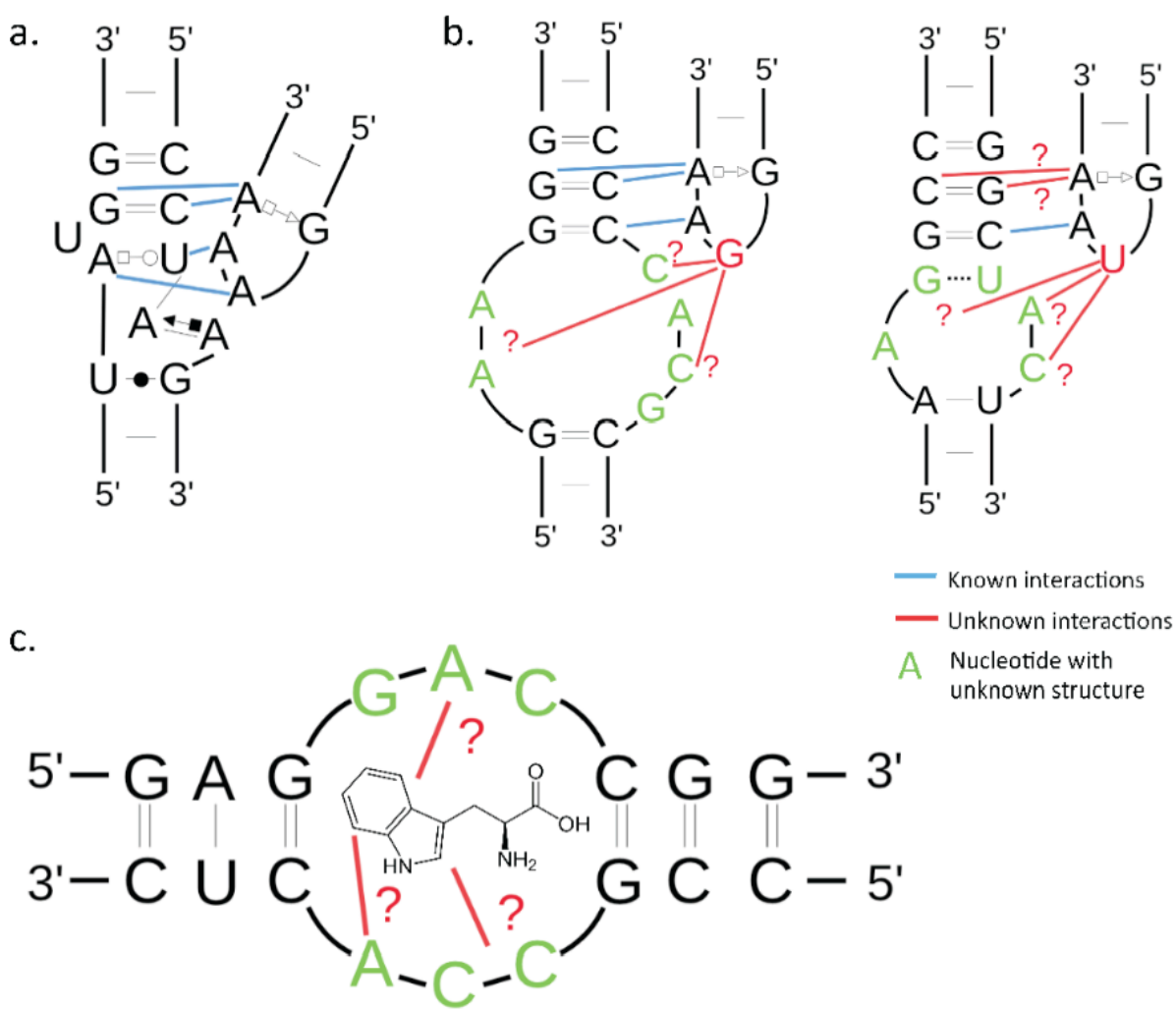